# Facilitating field-free perpendicular magnetization switching with a Berry curvature dipole in a Weyl semimetal

Dong Li [1,2,*], Xing-Yu Liu [1,*], Xing-Guo Ye,[1] Zhen-Cun Pan,[1] Wen-Zheng Xu,[1] Peng-Fei Zhu,[1] An-Qi Wang,[1]
Kenji Watanabe,[3] Takashi Taniguchi,[4] and Zhi-Min Liao [1,5,†]

[1]*State Key Laboratory for Mesoscopic Physics and Frontiers Science Center for Nano-optoelectronics, School of Physics,
Peking University, Beijing 100871, China*
[2]*Academy for Advanced Interdisciplinary Studies, Peking University, Beijing 100871, China*
[3]*Research Center for Electronic and Optical Materials, National Institute for Materials Science, 1-1 Namiki, Tsukuba 305-0044, Japan*
[4]*Research Center for Materials Nanoarchitectonics, National Institute for Materials Science, 1-1 Namiki, Tsukuba 305-0044, Japan*
[5]*Hefei National Laboratory, Hefei 230088, China*

We report the synergy between orbital and spin-orbit torques in $WTe_2/Fe_3GeTe_2$ heterostructures characterized by a Berry curvature dipole. By applying a current along the $a$ axis in $WTe_2$, we detect an out-of-plane magnetization in the system, which we attribute to nonequilibrium orbital magnetization linked to the Berry curvature dipole based on first-principles calculations, manifesting as the orbital Edelstein effect. This effect generates orbital torques that enable field-free perpendicular magnetization switching. Furthermore, by applying a relatively small current along the $a$ axis and a pulsed current along the $b$ axis in $WTe_2$, we demonstrate controllable field-free magnetization switching of the adjacent $Fe_3GeTe_2$ layer, independently manipulating the orbital and spin-orbit torques. Our findings not only enhance the understanding of the collaborative dynamics between these torques but also suggest potential applications in magnetoresistive random-access memory.

Nonvolatile magnetoresistive random-access memory with perpendicular magnetization (PM) is promising for the next-generation memory and logic applications [1–5]. Information is encoded by electrically manipulating the magnetic states of free layers, conventionally utilizing the mechanisms of spin-transfer torque [6] or spin-orbit torque (SOT) [7–10], whereas, with the increasing challenges of energy consumption, durability, and compatibility, the achievement of deterministic field-free magnetization switching of PM is essential [11–15]. Owing to symmetry restrictions of typical SOT, additional artificial engineering is required for field-free mechanisms, such as the additional interlayer exchange coupling [11,12] and in-plane structural and composition asymmetry [13,14]. Instead, the orbital angular momentum and orbital magnetic moment have the nature of the out-of-plane direction in two-dimensional systems [16,17], favoring the utilization for PM switching.

The emergent van deer Waals (vdW) materials, especially the topological semimetals, provide a unique opportunity for orbital physics [18–21]. The orbital magnetic moment $m_{orb}$ manifests as the self-rotation of Bloch electron wave packets and correlates with Berry curvature $\mathbf{\Omega}$. For instance, $m_{orb}$ calculated in the gapped Dirac system is 30 times larger than the Bohr magneton, due to the giant Berry curvature [16]. In type-II Weyl semimetals with low-crystalline symmetry, such as $WTe_2$ and $TaIrTe_4$, the Berry curvature dipole (BCD) $\mathbf{D}$ as the intrinsic origin of nonlinear Hall effect (NLHE) has been demonstrated, having a dipolelike distribution of $\mathbf{\Omega}$ along the $a$ axis [22–26]. Thus, $m_{orb}$ exhibits a similar dipole-type texture. With an electric field applied parallel to $\mathbf{D}$ [along the $a$ axis in Fig. 1(a)], an out-of-plane orbital magnetization can be induced, known as the orbital Edelstein effect [27–31], facilitating the field-free magnetization switching in adjacent magnetic layers [32–35]. Moreover, owing to the strong spin-orbit coupling and nontrivial band structures in $WTe_2$ and $TaIrTe_4$ [36–40], when applying a current along the $b$ axis, a large spin polarization and significant SOT are generated for in-plane magnetization switching [37]. Although the field-free PM switching has been realized, facilitated by the out-of-plane antidampinglike torque [34,35,37–40], the orbital contribution related to BCD remains ambiguous, and the synergy dynamics between SOT and orbital torques are still elusive.

Here we investigate the synergy between orbital and spin-orbit torques in $WTe_2/Fe_3GeTe_2$ (FGT) heterostructures. Utilizing a magnetic detector electrode, we observe a current-induced out-of-plane magnetization in $WTe_2$. Based on first-principles calculations, we reveal the underlying current-induced orbital magnetization related to BCD, identifying it as a major contribution of the out-of-plane magnetization. We further introduce a different approach via two distinct driving currents, a pulsed current $I_p$ along the $b$ axis and a relatively small DC current $I_{DC}$ along the $a$ axis of $WTe_2$, to independently manipulate the spin orbit and orbital torques, respectively. The collaborative dynamics between the two torques are demonstrated with support from micromagnetic simulations, showcasing the determinative role of the orbital torques in the field-free magnetization switching process.

[*]These authors contributed equally to this work.
[†]Contact author: liaozm@pku.edu.cn

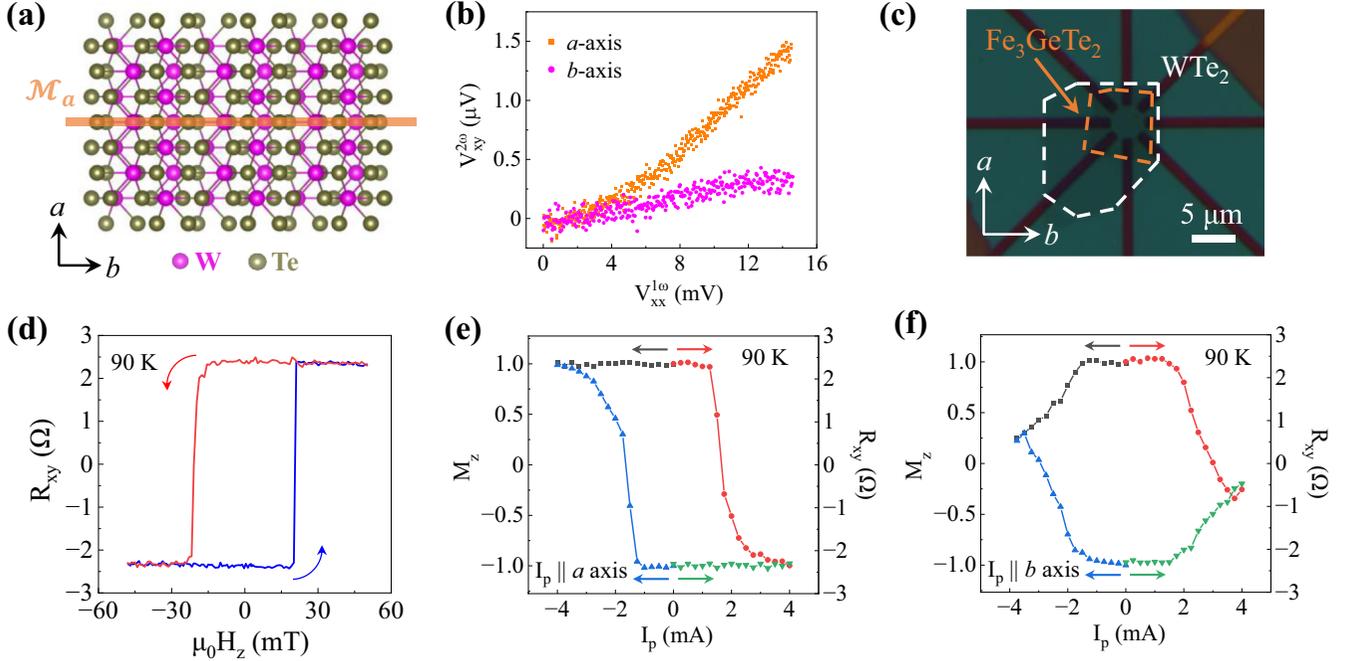

FIG. 1. (a) Crystal structure of few-layer WTe$_2$ from top views. The yellow solid line represents the mirror line $\mathcal{M}_a$. (b) The NLHE measured in a few layer WTe$_2$ flake at 1.6 K with an AC current $I^\omega$ along the $a$ axis and $b$ axis, respectively. The $V_{xx}^{1\omega} = I^\omega R_{xx}$, where $R_{xx}$ is the longitudinal resistance. (c) Optical image of device A. Fe$_3$GeTe$_2$ (6 nm) and WTe$_2$ (4.2 nm) are distinctly delineated by orange and white dashed borders, respectively. (d) Anomalous Hall resistance in device A at 90 K. (e),(f) Pulsed current $I_p$ induced magnetization switching at 90 K, reflected by the anomalous Hall resistance, for $I_p$ applied along the (e) $a$ axis and (f) $b$ axis.

$T_d$-WTe$_2$ is a type-II Weyl semimetal with broken inversion symmetry [41]. In contrast to the bulk, twofold rotational symmetry is broken at the surface [Fig. 1(a)], having a nonzero BCD along its low-symmetry axis ($a$ axis) [23]. NLHE [23–26] is observed in a few-layer WTe$_2$ device when an AC current $I^\omega$ is applied along the $a$ axis [Fig. 1(b) and Fig. S1 of the Supplemental Material [52]]. Figure 1(c) shows the optical image of WTe$_2$/FGT heterostructure device A. By identifying long, straight edges of WTe$_2$ and combining with polarized Raman spectroscopy, the crystalline axes are aligned with the electrodes. The layered vdW ferromagnetic (FM) metal FGT has strong perpendicular magnetic anisotropy (PMA) [42,43]. By sweeping an out-of-plane magnetic field, the anomalous Hall resistance $R_{xy}$ exhibits a square-shaped hysteresis loop [Fig. 1(d)]. The Curie temperature of FGT in device A is approximately 150 K [Fig. S2(c) [52]]. We then examine the magnetization switching by applying a pulsed current $I_p$ and monitoring $R_{xy}$ with a small AC current. Notably, when $I_p$ is applied along the $a$ axis, a full PM switching with a critical current density about $8.5 \times 10^6$ A/cm$^2$ is observed under zero external magnetic field [Fig. 1(e)], indicating the presence of out-of-plane antidampinglike torques. Furthermore, by integrating a WTe$_2$ thin layer with the FGT/$h$-BN/FGT magnetic tunnel junction, all-electric control of van der Waals magnetoresistive memory has been achieved (Fig. S4 [52]). In contrast, when sweeping $I_p$ from 0 to ±4 mA along the $b$ axis, nearly demagnetized states ($R_{xy} \sim 0$) are observed with only in-plane SOT [Fig. 1(f)].

To explore more understanding of the field-free PM switching, we perform the measurement of the current-induced out-of-plane magnetization in WTe$_2$ using the measurement configuration depicted in Fig. 2(a), where a FGT flake is used as the FM probe, and a $h$-BN thin layer is stacked between WTe$_2$ and FGT, which acts as a tunneling barrier to improve the detection efficiency. When applying a fixed current $I^\omega$ in WTe$_2$, the out-of-plane magnetization in WTe$_2$ ($M_{\text{WTe}_2}$) is measured via a voltage difference ($V_m^\omega$) between the FGT electrode and a reference Au electrode. An out-of-plane magnetic field is applied to manipulate the magnetization of the FGT, $M_{\text{FGT}}$. When $M_{\text{WTe}_2}$ is parallel to $M_{\text{FGT}}$, i.e., $M_{\text{WTe}_2} \parallel M_{\text{FGT}}$, a low voltage state will be detected. Conversely, $M_{\text{WTe}_2} \parallel -M_{\text{FGT}}$ will produce a measurable high voltage state. The mechanism of magnetization detection essentially involves the measurement of magnetic moment-dependent chemical potentials, a method that has been widely applied in the study of topological materials [44–46].

We demonstrate the results of device B. The optical image of device B is shown in Fig. 2(b), in which we can perform angle-dependent measurements by rotating the current direction. When applying a current along the $a$ axis in WTe$_2$, as shown in Fig. 2(c), the hysteresis behavior of $V_m^\omega$ upon sweeping the magnetic field is observed ($V_m^\omega$ is obtained by subtracting the background in the raw data). The difference between high and low states of $V_m^\omega$ indicates a nonzero $M_{\text{WTe}_2}$ in WTe$_2$. However, the hysteresis loop is hardly visible when applying the current along the $b$ axis, as shown in Fig. 2(d). Furthermore, using two nonmagnetic electrodes as the voltage probes, no hysteresis loop is observed [Fig. S5(a) [52]], which rules out the magnetization-dependent signal

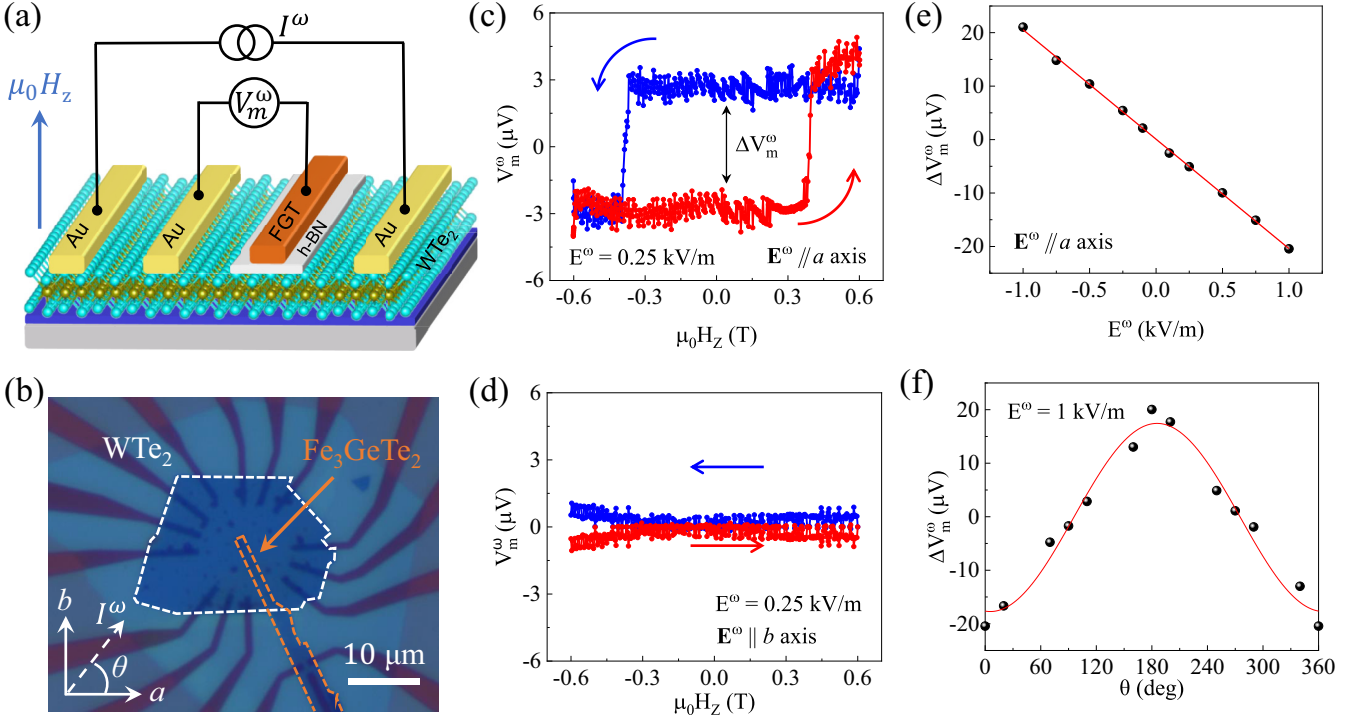

FIG. 2. (a) Schematic of magnetization detection in WTe$_2$ via a FGT electrode. (b) Optical image of device B. Fe$_3$GeTe$_2$ and WTe$_2$ are distinctly delineated by orange and white dashed borders, respectively. (c),(d) The magnetization-dependent voltage $V_m^\omega$ when applying the driving current $I^\omega$ along the (c) $a$ axis and (d) $b$ axis with $E^\omega = 0.25$ kV/m. The $E^\omega = I^\omega R_{xx}/L_{xx}$, where $L_{xx}$ is the length of the channel. (e) The loop height $\Delta V_m^\omega$ as a function of $E^\omega$ when applying the driving current along the $a$ axis. (f) $\Delta V_m$ as a function of angle $\theta$ under $E^\omega = 1$ kV/m, where $\theta$ is defined as the angle between the $a$ axis and the applied current. All data were measured at 5 K.

originating from the proximity induced magnetism in WTe$_2$ by FGT.

We further investigated the $V_m^\omega$ by varying the magnitude of the driving current along the $a$ axis [Fig. S5(b) [52]]. The loop height ($\Delta V_m^\omega$) is extracted as the voltage difference at zero magnetic field between forward and backward sweeping magnetic field, which is proportional to the magnetization $M_{WTe_2}$ in WTe$_2$. A linear dependence of $\Delta V_m^\omega$ is found on the driving current [Fig. 2(e)], indicating that the $M_{WTe_2}$ is proportional to the driving current $I^\omega$. By exploiting the multielectrode pairs of device B, the $M_{WTe_2}$ is studied as $I^\omega$ is applied along different directions. The $\Delta V_m^\omega$ is measured under different angle $\theta$, where $\theta$ is defined as the angle between $I^\omega$ and the $a$ axis [inset in Fig. 2(b)]. As shown in Fig. 2(f), the $\Delta V_m^\omega$ shows a cosine dependence on $\theta$ with a maximum value along the $a$ axis and a negligible value along the $b$ axis. The angle-dependent behavior is quite consistent with the PM switching results in Fig. 1 and Fig. S2 [52]. We conclude that the current-induced out-of-plane magnetization is an origin of the out-of-plane antidampinglike torque.

To reveal the origin of the current-induced out-of-plane magnetization, we perform first-principles calculations with a few-layer WTe$_2$ slab model (see Supplemental Material Note 7 [52]). Figure 3(a) displays the calculated band structure along high-symmetry directions in the Brillouin zone. The semimetallic feature is identified from the overlapping electron and hole pockets along the $k$ path from $\Gamma$ to $X$. The out-of-plane components of Berry curvature and orbital magnetic moment distributions in $k$ space at the calculated Fermi energy ($\mu = 0$ eV) are plotted in Figs. 3(b) and 3(c), respectively. Notably, the Berry curvature exhibits $\Omega_z(k_x, k_y) = -\Omega_z(-k_x, k_y)$ due to the mirror symmetry $\mathcal{M}_a$, indicating a finite BCD component $D_{xz}$. Owing to the close correlation between the Berry curvature and orbital magnetic moment, the orbital magnetic moment $m_{orb}^z$ shows an analogous distribution with $\Omega_z$. The $m_{orb}^z$ is particularly large and even reaches about $16\mu_b$ (where $\mu_b$ denotes the Bohr magneton) around the hotspots of Berry curvature. Moreover, the dipole-type texture of $m_{orb}^z$ suggests a current-induced out-of-plane orbital magnetization when applying a driving current along the $a$ axis of WTe$_2$.

According to the orbital Edelstein effect theories [27–29,47], the electric field $\mathbf{E}$ induced orbital magnetization $\mathbf{M_{orb}}$ can be described as $M_{orb}^j = \sum_{i,j} \frac{e^2\tau}{2\hbar^2}\alpha_{ij}E_i$, where $i,j = x,y,z$, $\tau$ is the relaxation time, and $\alpha_{ij}$ is orbital Edelstein coefficients. The BCD could contribute to the orbital Edelstein coefficients $\alpha_{ij}$, formulated as $\alpha_{ij} = -\mu D_{ij} + B_{ij}$, where $\mu$ is the chemical potential, and $-\mu D_{ij}$ and $B_{ij}$ represent the contributions of BCD and effective magnetic field, respectively [47]. The calculated orbital Edelstein coefficient $\alpha_{xz}$ of WTe$_2$ and its contribution from BCD are depicted in Fig. 3(d). Based on the experimentally measured carrier densities, we estimate the Fermi level of WTe$_2$ at 100 K corresponding to $\mu_F = 63 \pm 5$ meV in our first-principles tight-binding Hamiltonian (see Supplemental Material Note 8 [52]). It can be seen that the coefficient $\alpha_{xz}$ is close to $-\mu D_{xz}$ around the experimental Fermi energy [denoted by the shaded range in Fig. 3(d)], indicating the dominant contribution of BCD to the orbital magnetization. Moreover, we compare the spin and orbital Edelstein effects in WTe$_2$ (Supplemental Material Note 7 [52]). As shown in Fig. 3(e), under an electric field $E_x = 10^5$ V/m along the $a$ axis of WTe$_2$, the obtained out-of-plane orbital magnetization reaches 0.1 $\mu_b$/nm$^2$, which is two orders of magnitude larger than the spin magnetization at the Fermi level. This finding supports the orbital origin of the antidampinglike torque and highlights the critical role of orbital magnetization in field-free PM switching.

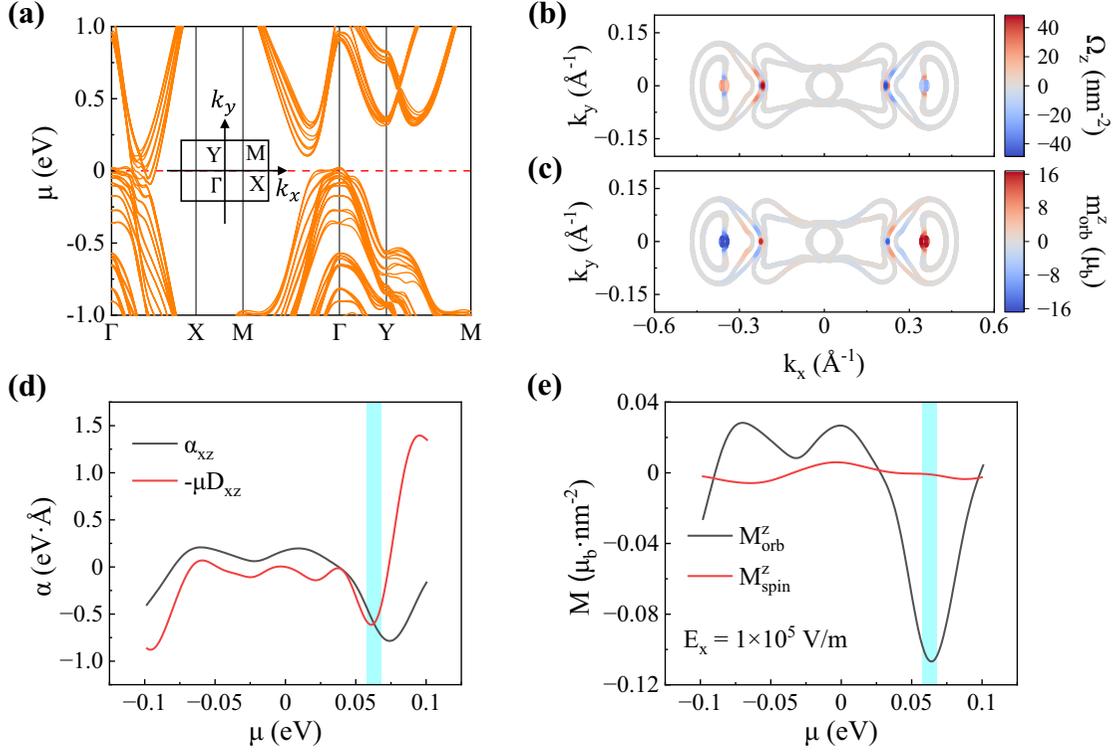

FIG. 3. (a) Energy band structure of a five-layer WTe$_2$ obtained by *ab initio* calculations. (b),(c) Calculated momentum $k$-resolved distribution of the $z$ component of (b) Berry curvature and (c) orbital magnetic moment at energy levels $\mu = 0$ eV [indicated by the red dashed line in (a)]. (d) Calculated orbital magnetoelectric susceptibility $\alpha_{xz}$ and its BCD contribution $(-\mu D_{xz})$ as a function of chemical potential $\mu$. The shaded range indicates the estimated Fermi level of WTe$_2$ at 100 K, with a deviation of $\pm 5$ meV. The coordinates $x$, $y$, $z$ correspond to the $a$, $b$, and $c$ axes of WTe$_2$, respectively. (e) Comparison between the out-of-plane orbital and spin magnetization as a function of chemical potential $\mu$, assuming an external electric field of $10^5$ V/m applied along the $a$ axis of WTe$_2$. A temperature of 100 K and a relaxation time of 1 ps are used for the calculations.

We further utilize orbital torques in synergy with spin-orbit torques to demonstrate efficient field-free magnetization switching in WTe$_2$/FGT heterostructures. The measurement configuration is illustrated in Fig. 4(a) (see Supplemental Material Note 9 [52] for circuit schematic), where two distinct driving currents are used to independently manipulate the magnitude of the two torques. We apply a DC current $I_{DC}$ along the $a$ axis, a pulsed current $I_p$, and an ultralow AC current (1 μA) along the $b$ axis. $I_p$ and $I_{DC}$ are applied to generate SOT and orbital torques, respectively [Fig. 4(b)], while the AC current is applied to monitor the anomalous Hall resistance $R_{xy}$ after $I_p$ is removed. Both torques induce the precession of magnetic moments in FGT until stabilization at final sates. As shown in Fig. 4(c), when $I_{DC} = +0.4$ mA is applied, partial switching (about 65%) is observed, and the final net magnetization state $M_z^{\text{Final}}$ of FGT points downward regardless of whether $I_p$ is positive or negative. Conversely, when $I_{DC}$ is $-0.4$ mA [Fig. 4(d)], $M_z^{\text{Final}}$ points upward. Notably, the applied $I_{DC}$ is one order of magnitude smaller than the critical switching current of $I_p$. It suggests that small out-of-plane antidampinglike torques can act to break the symmetry and achieve deterministic PM switching when combined with in-plane SOT. Note that similar results are reproducible for another device (Fig. S9 [52]). It is found that the partial switching is strongly dependent on the magnitude of $I_{DC}$, as depicted in Fig. 4(e), where the final net magnetization $|M_z^{\text{Final}}|$ gradually approaches 1, showing that larger out-of-plane antidampinglike torques lead to more complete switching of multiple domains in FGT. In addition, each magnitude of $I_{DC}$ corresponds to a specific $M_z^{\text{Final}}$, suggesting the potential for developing multilevel nonvolatile memory devices with applications in neuromorphic and in-memory computing [48–50].

We also perform the current induced exchange bias in device A. As depicted in Fig. 4(f), with $I_p = +3.5$ mA along the $b$ axis, the center of anomalous Hall effect (AHE) loop shifts towards positive (negative) magnetic fields when $I_{DC} = +0.3$ mA ($I_{DC} = -0.3$ mA). Note that the direction of AHE loop shifts is independent of the polarity of $I_p$ along the $b$ axis [Fig. 4(f) and Fig. S10(c) [52]], consistent with the observed switching results in Figs. 4(c) and 4(d). However, no obvious

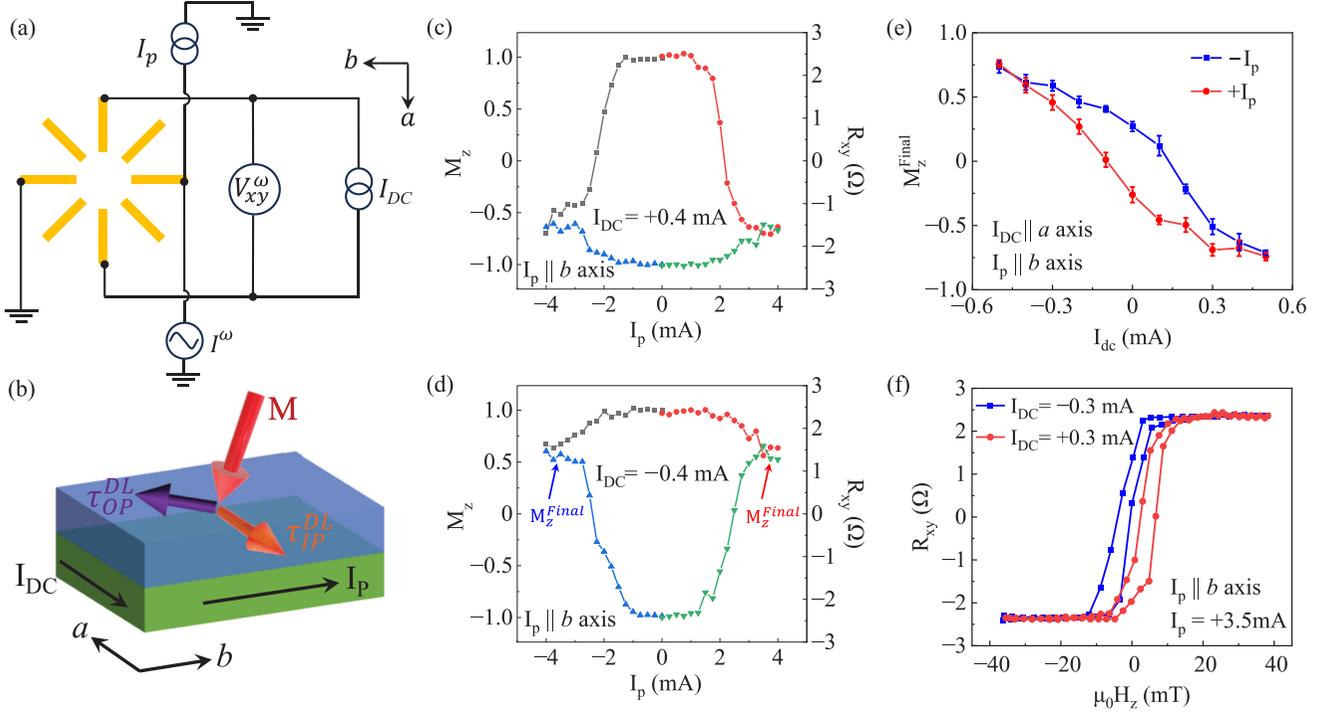

FIG. 4. (a) Illustration of measurement configuration. (b) Schematic depiction of the $I_p$ and $I_{DC}$ induced the in-plane ($\tau_{IP}^{DL}$) and out-of-plane ($\tau_{OP}^{DL}$) antidampinglike torques, respectively. (c),(d) The pulsed current $I_p$ along the $b$ axis induced magnetization switching for the DC current $I_{DC}$: (c) $+0.4$ mA and (d) $-0.4$ mA along the $a$ axis. (e) The final net magnetization ($M_z^{final}$) as a function of $I_{DC}$ for the different current combinations of $I_p$ and $I_{DC}$. (f) AHE hysteresis loops measured at 90 K with $I_p = +3.5$ mA along the $b$ axis and $I_{DC} = \pm 0.3$ mA along the $a$ axis.

AHE loop shift is observed when $I_p = \pm 3.5$ mA is applied along the $b$ axis in the absence of $I_{DC}$ or when $I_{DC} = \pm 0.3$ mA is applied along the $a$ axis alone in the absence of $I_p$ [Figs. S10(b) and S11(b) [52]], which suggests the synergistic effect of SOT and orbital torques in field-free PM switching. The collaborative dynamics between the SOT and orbital torques are further revealed by micromagnetic simulations [51] (see Supplemental Material Note 13 [52]), which align well with the experimental findings, underscoring the determinative role of the orbital torque in the field-free PM switching process.

In summary, we have demonstrated the synergy between orbital and spin-orbit torques in WTe$_2$/Fe$_3$GeTe$_2$ heterostructures. By applying two distinct driving currents, we independently manipulate these torques, demonstrating controlled full and partial PM switching. Our findings clarify that current-induced orbital magnetization, associated with the BCD, plays a significant role in generating the out-of-plane antidampinglike torque, as evidenced by magnetization detection experiments and first-principles calculations. Our work provides deeper insights into electrically controlled field-free PM switching, paving the way for device applications harnessing orbital effects and Berry curvature dipoles.

This work was supported by the National Natural Science Foundation of China (Grants No. 62425401 and No. 62321004) and the Innovation Program for Quantum Science and Technology (Grant No. 2021ZD0302403). K.W. and T.T. acknowledge support from the JSPS KAKENHI (Grants No. 20H00354, No. 21H05233, and No. 23H02052) and World Premier International Research Center Initiative (WPI), MEXT, Japan.

# Supplemental Material for

# Facilitating field-free perpendicular magnetization switching with Berry curvature dipole in Weyl semimetals

Dong Li[1,2,+], Xing-Yu Liu[1,+], Xing-Guo Ye[1], Zhen-Cun Pan[1], Wen-Zheng Xu[1], Peng-Fei Zhu[1], An-Qi Wang[1], Kenji Watanabe[3], Takashi Taniguchi[4], Zhi-Min Liao[1,5*]

[1]State Key Laboratory for Mesoscopic Physics and Frontiers Science Center for Nano-optoelectronics, School of Physics, Peking University, Beijing 100871, China.

[2]Academy for Advanced Interdisciplinary Studies, Peking University, Beijing 100871, China.

[3]Research Center for Electronic and Optical Materials, National Institute for Materials Science, 1-1 Namiki, Tsukuba 305-0044, Japan.

[4]Research Center for Materials Nanoarchitectonics, National Institute for Materials Science, 1-1 Namiki, Tsukuba 305-0044, Japan

[5]Hefei National Laboratory, Hefei 230088, China.

+ These authors contributed equally.

* Email: liaozm@pku.edu.cn

**This file contains Supplemental Figures S1-S14 and Notes 1-13.**

**Note 1:** Device fabrication and measurement methods.

**Note 2:** Nonlinear Hall effect in few layer $WTe_2$ device.

**Note 3:** Additional data of Device A.

**Note 4:** Critical current density with the parallel current model.

**Note 5:** All van der Waals magnetoresistive memory device.

**Note 6:** Additional data of the current-induced magnetization in $WTe_2$.

**Note 7:** Theoretical calculations of orbital and spin Edelstein coefficients.

**Note 8:** Experimental and calculated carrier densities of few-layer $WTe_2$.

**Note 9:** Circuit schematic with multiple sources in experiments.

**Note 10:** Reproducible results in Device C.

**Note 11:** AHE hysteresis loop shift measurements in Device A.

**Note 12:** Discussion of the heating effect in field-free perpendicular magnetization switching.

**Note 13:** Micromagnetic simulations.

**Supplemental Note 1: Device fabrication and measurement methods.**

**(1) Device fabrication**

WTe$_2$, h-BN and Fe$_3$GeTe$_2$ flakes were prepared though mechanical exfoliation onto SiO$_2$/Si substrates, and thin flakes were identified by optical contrast. The van der Waals heterostructures were fabricated utilizing the dry transfer technique. For the WTe$_2$/Fe$_3$GeTe$_2$ heterostructure devices, the layers of h-BN, Fe$_3$GeTe$_2$, and WTe$_2$ were picked up using polycarbonate (PC) film and placed onto prepatterned Ti/Au electrodes (~12 nm thick) on individual SiO$_2$/Si substrates. For the magnetoresistive memory device, the flakes were sequentially picked up in the following order: capping h-BN, top Fe$_3$GeTe$_2$, thin h-BN, bottom Fe$_3$GeTe$_2$, and WTe$_2$. The entire exfoliation and transfer processes were conducted in an argon-filled glove box with O$_2$ and H$_2$O content below 0.01 parts per million to avoid sample degradation.

**(2) Raman measurements**

The Raman spectroscopy was conducted using a 532 nm excitation wavelength through a linearly polarized solid-state laser beam. The scattered Raman signals were collected through the objective lens and directed to a spectrometer. Polarized Raman spectroscopy in the parallel configuration was utilized to determine crystalline axes of WTe$_2$.

**(3) Transport Measurements**

The devices were measured in an Oxford cryostat equipped with a variable temperature insert and a superconducting magnet. The first- and second-harmonic voltage signals were measured utilizing Stanford Research Systems SR830 and SR865A lock-in amplifiers with a typical frequency of ω = 17.777 Hz, unless otherwise specified. A Keithley 2400 current source, a Keithley 6221 current source and a Keithley 2182A nanovoltmeter were utilized for current-induced magnetization switching and AHE loop shift experiments. The used pulsed current $I_p$ was a square-wave pulse with varying magnitude and a width of 60 μs.

**Supplemental Note 2: Nonlinear Hall effect in few layer WTe₂ device.**

To detect the Berry curvature dipole in WTe$_2$, we fabricated a few-layer WTe$_2$ (~5 nm) device and measured the second-order nonlinear Hall effect (NLHE). Figure S1 shows the second-harmonic transverse voltage ($V_{xy}^{2\omega}$) for different driving frequencies. The harmonic current $I^\omega$ is applied along the *a*-axis of WTe$_2$. No frequency dependence is observed in the frequency range we have investigated (17.777–177.77 Hz). It validates the essential property of the BCD and excludes potential measurement artifacts such as spurious capacitive coupling.

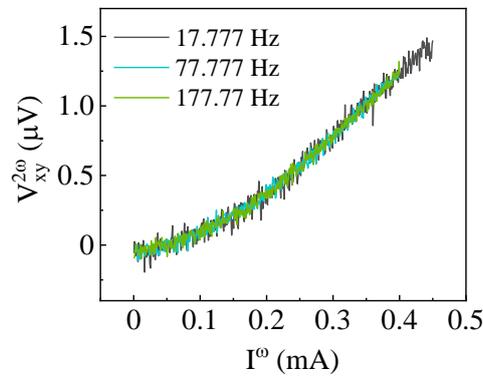

**Fig. S1.** The NLHE measured for different driving frequencies at 1.6 K. The harmonic current $I^\omega$ is applied along the *a*-axis of WTe$_2$.

# Supplemental Note 3: Additional data of Device A.

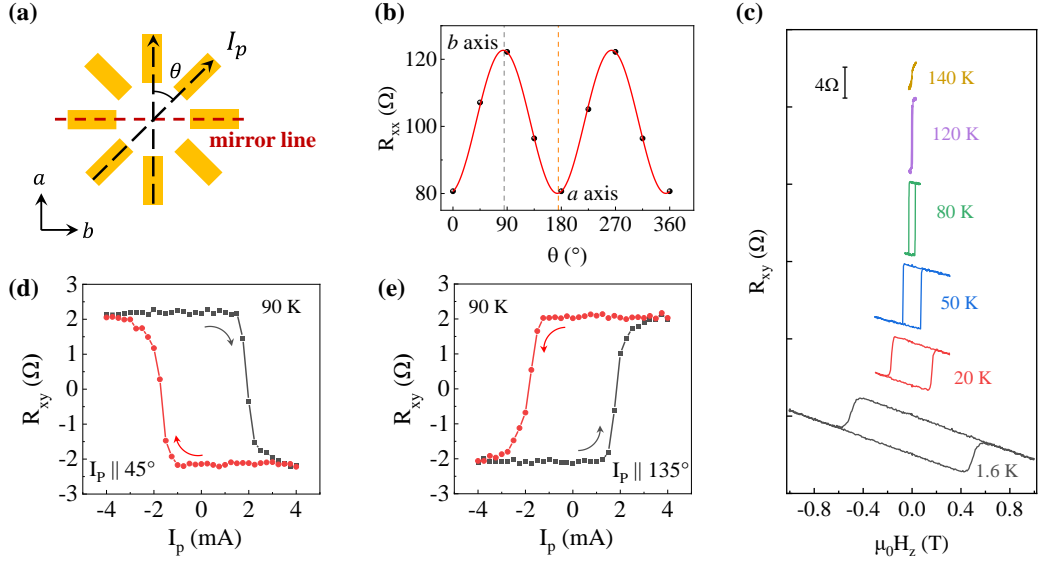

**FIG. S2. (a)** Schematic of the measurement configuration, where $\theta$ is the angle between $I_p$ and the $a$-axis.

**(b)** Longitudinal resistance anisotropy of Device A, where the resistance along the $b$-axis is larger than the $a$-axis of WTe$_2$ in Device A.

**(c)** The Hall resistance as a function of magnetic field at various temperatures in Device A.

**(d, e)** The Hall resistance as a function of pulse current $I_p$ along (d) $\theta = 45°$ and (e) $\theta = 135°$ at 90 K. The arrows show the (d) clockwise and (e) counterclockwise switching polarity.

**Supplemental Note 4: Critical current density with the parallel current model.**

For the Device A, the critical current is about 3 mA when $I_p$ is applied along the $a$-axis, as shown in Fig. 1(e). The resistivity $\rho_{xx}$ of Device A along the $a$-axis and a few-layer WTe$_2$ (4.6 nm) device along the $a$-axis as a function of temperature were measured, as shown in Fig. S3. By considering a parallel resistance model, we obtained the resistivity of FGT by $\rho_{xx}^{FGT} = \frac{t_{FGT}}{\left(\frac{t}{\rho_{xx}} - \frac{t_{WTe_2}}{\rho_{xx}^{WTe_2}}\right)}$, where $\rho_{xx}^{WTe_2}$ is the resistivity of WTe$_2$, $\rho_{xx}^{FGT}$ is the resistivity of Fe$_3$GeTe$_2$, $t_{WTe_2}$ is the thickness of WTe$_2$, $t_{FGT}$ is the thickness of Fe$_3$GeTe$_2$ and $t = t_{WTe2} + t_{FGT}$. For device A, the measured thicknesses $t_{WTe_2} = 4.2$ nm and $t_{FGT} = 6$ nm. Furthermore, the current distribution in the WTe$_2$ layer is estimated by $\frac{I_{WTe_2}}{I} = \frac{1}{1 + \frac{\rho_{xx}^{WTe_2} t_{FGT}}{\rho_{xx}^{FGT} t_{WTe_2}}}$, where $I$ is the applied current flowing in the whole heterostructure, and $I_{WTe_2}$ is the current component flowing in the WTe$_2$ layer. According to the above formula, we estimated the current distribution in the WTe$_2$ layer ($I_{WTe_2}/I$) is about 0.48 and further obtained a critical current density about $8.5 \times 10^6$ A/cm$^2$ when $I_p$ is applied along the $a$-axis in Device A at 90 K.

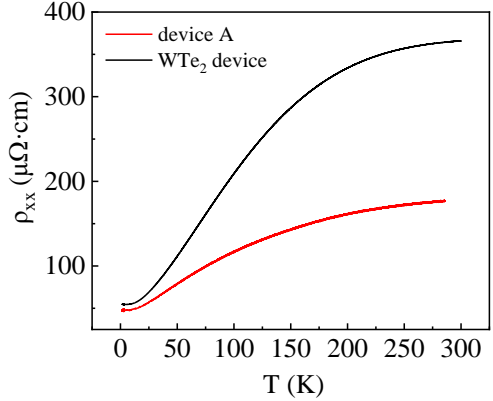

**FIG. S3.** The resistivity as a function of temperature for Device A and a few layer WTe$_2$ device.

**Supplemental Note 5: All van der Waals magnetoresistive memory device.**

Leveraging the achievement of field-free magnetization switching in the WTe$_2$/FGT heterostructure, we further combined the WTe$_2$ layer with an FGT/thin h-BN/FGT magnetic tunneling junction (MTJ) to demonstrate a vdW memory device, as schematically show in Fig. S4(a). High-quality and atomically thin h-BN serves as a tunneling barrier, ensuring dependable data reading via the tunneling magnetoresistance (TMR) effect [58]. The states of "1" and "0" correspond to the high and low resistance states of the tunnel junction, respectively, depending on the antiparallel and parallel magnetizations of the two FGT layers, as shown in Figs. S4(b) and S4(c). The magnetoresistive memory's data writing is accomplished by field-free switching the magnetization of the bottom FGT layer through the application of current pulses along the *a*-axis, as shown in Fig. S4(d). Utilizing a series of ±1.25 mA writing current pulses, the state transition between "0" and "1" demonstrates robustness and non-volatility [Fig. S4(e)]. Besides, the extracted critical current density and TMR ratio [Fig. S4(f)] decrease with increasing temperature, consistent with the reduction of FGT ferromagnetism. This memory shows a TMR ratio about 10% at 1.6 K and a switching current density ~$10^6$ A·cm$^{-2}$ at 150 K.

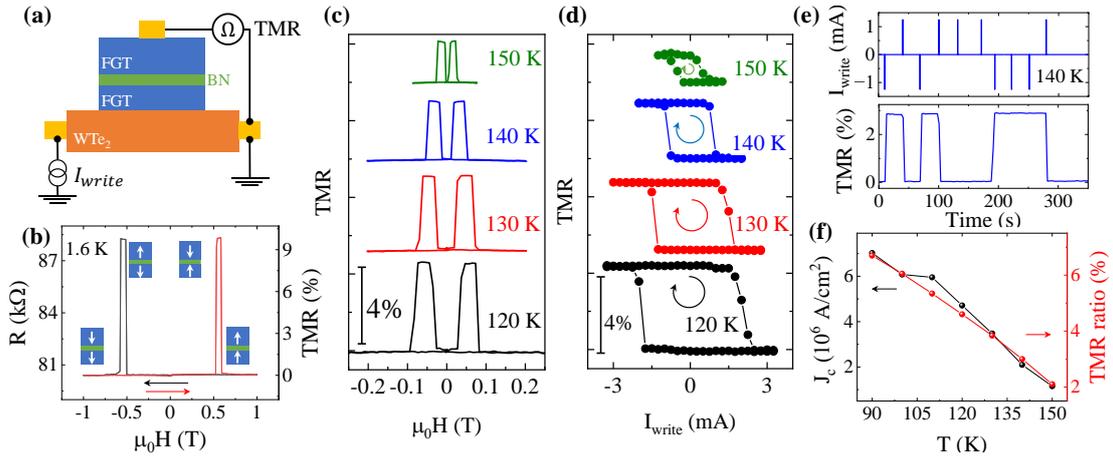

**FIG. S4. (a)** Schematic of the magnetoresistive memory. The data is recorded by applying a write current, $I_{write}$, to the WTe$_2$ channel. The output is retrieved through the TMR of the tunneling junction.
**(b)** The TMR as a function of magnetic field at 1.6 K. The inset shows the magnetization of the two FGT layers.

**(c, d)** The TMR as a function of (c) magnetic field and (d) $I_{write}$ at different temperatures. The circle arrows show the clockwise switching polarity.

**(e)** Upper panel: applying a sequence of ±1 mA pulsed currents for data writing at 140 K; lower panel: the corresponding data reading by the TMR values.

**(f)** The critical switching current density (black) and TMR ratio (red) as a function of temperature.

**Supplemental Note 6: Additional data of the current-induced magnetization in WTe$_2$.**

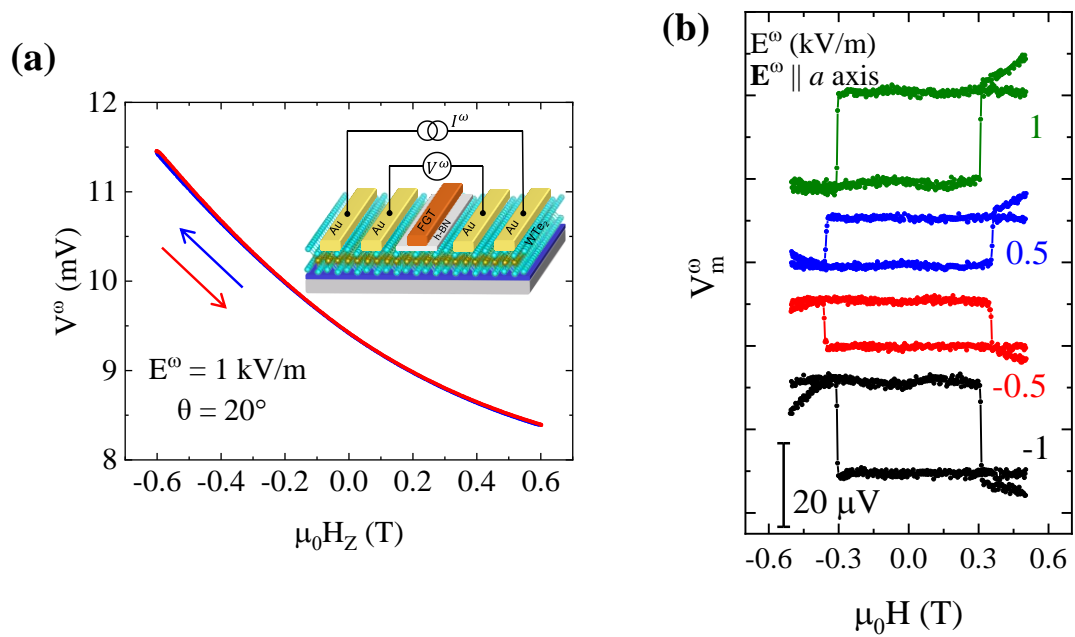

**FIG. S5. (a)** The voltage difference between two Au reference electrodes as a function of the magnetic field. No hysteresis loop is observed.

**(b)** $V_m^\omega$ under various $E^\omega$ using the FGT as the voltage probe when $E^\omega$ is applied along the *a*-axis. The loop height increases with excitation current and that the polarity of the loop reverses from counterclockwise to clockwise hysteresis when the excitation current is reversed (exchanging the source-drain leads). All data were measured at 5 K.

**Supplemental Note 7: Theoretical calculations of orbital and spin Edelstein coefficients.**

**(1) First-principles calculations**

We performed *ab initio* density-functional theory (DFT) calculations on a five-layer $T_d$-WTe$_2$ slab model and then constructed a DFT-based tight-binding model Hamilton. The tight-binding model matrix elements were calculated by projecting onto the Wannier orbitals [53,54]. Specifically, the *d* orbitals of W atoms and *p* orbitals of Te atoms were utilized to construct Wannier functions using Wannier90 code [55], without undergoing the procedure for maximizing localization. The Berry curvature dipole and orbital (spin) Edelstein coefficients were calculated using the WannierBerri code [55], with integration over the 2-dimensional Brillouin zone taken on a 1501 × 801 fine grid.

**(2) Orbital Edelstein coefficients**

The current-induced nonequilibrium orbital magnetization can be written as $\boldsymbol{M}_{\text{orb}} = \left(\frac{e^2\tau}{2\hbar^2}\right)\alpha_{\text{orb}}\boldsymbol{E}$ under an applied electric field $\boldsymbol{E}$. Owing to the few layer WTe$_2$ holding a mirror symmetry $\mathcal{M}_a$, the coefficient $\alpha_{\text{xz}}^{\text{orb}}$ is allowed, whereas $\alpha_{\text{yz}}^{\text{orb}}$ is constrained to be zero. According to the reported theory [47], there are two terms contribute to $\alpha_{\text{xz}}^{\text{orb}}$: one is the Berry curvature dipole contribution and the other is an effective magnetic field contribution. It is derived as follows:

$$\alpha_{\text{xz}} = \int_{BZ}\frac{d^2k}{(2\pi)^2}\sum_n m_{n\mathbf{k}}^{\text{z}}v_{n\mathbf{k}}^{\text{x}}\partial_{\epsilon_{n\mathbf{k}}}f_{n\mathbf{k}}^{(0)}$$

$$= i\int_{BZ}\frac{d^2k}{(2\pi)^2}\sum_{m\neq n,n}(\epsilon_{n\mathbf{k}} - \epsilon_{m\mathbf{k}})v_{n\mathbf{k}}^{\text{x}}\left(\frac{\langle u_{n\mathbf{k}}|\hat{v}_{\mathbf{k}}|u_{m\mathbf{k}}\rangle \times \langle u_{m\mathbf{k}}|\hat{v}_{\mathbf{k}}|u_{n\mathbf{k}}\rangle}{(\epsilon_{n\mathbf{k}} - \epsilon_{m\mathbf{k}})^2}\right)_z \partial_{\epsilon_{n\mathbf{k}}}f_{n\mathbf{k}}^{(0)}$$

$$= \int_{BZ}\frac{d^2k}{(2\pi)^2}\sum_n(\epsilon_{n\mathbf{k}}\Omega_{n\mathbf{k}}^{\text{z}} - b_{n\mathbf{k}}^{\text{z}})v_{n\mathbf{k}}^{\text{x}}\partial_{\epsilon_{n\mathbf{k}}}f_{n\mathbf{k}}^{(0)}$$

$$= -\mu D_{\text{xz}} + B_{\text{xz}}, \quad (S1)$$

where the Berry curvature dipole $D_{xz}$ is calculated as [22]

$$D_{\text{xz}} = -\int_{BZ}\frac{d^2k}{(2\pi)^2}\sum_n \Omega_{n\mathbf{k}}^{\text{z}}v_{n\mathbf{k}}^{\text{x}}\partial_{\epsilon_{n\mathbf{k}}}f_{n\mathbf{k}}^{(0)}, \quad (S2)$$

and the effective magnetic field term related to the band structure is defined as [47]

$$\boldsymbol{b}_{n\mathbf{k}}(\omega) = i\sum_{m\neq n}\epsilon_{m\mathbf{k}}\frac{\langle u_{n\mathbf{k}}|\hat{v}_{\mathbf{k}}|u_{m\mathbf{k}}\rangle \times \langle u_{m\mathbf{k}}|\hat{v}_{\mathbf{k}}|u_{n\mathbf{k}}\rangle}{(\epsilon_{n\mathbf{k}}-\epsilon_{m\mathbf{k}})^2-(\hbar\omega+i\Gamma)^2}. \quad (S3)$$

In the above equations, $m_{n\mathbf{k}}^z$ is out-of-plane component of the orbital magnetic moment and defined as [17]

$$m_{n\mathbf{k}}^z = i \sum_{m \neq n} \left( \frac{\langle u_{n\mathbf{k}}|\hat{v}_{\mathbf{k}}|u_{m\mathbf{k}}\rangle \times \langle u_{m\mathbf{k}}|\hat{v}_{\mathbf{k}}|u_{n\mathbf{k}}\rangle}{(\epsilon_{n\mathbf{k}} - \epsilon_{m\mathbf{k}})} \right)_z, \quad (S4)$$

where $v_{n\mathbf{k}}^x$ is the band velocity along the *a*-axis, $\mu$ is the chemical potential, and $\Omega_{n\mathbf{k}}^z$ is the out-of-plane component of Berry curvature.

**(3) Spin Edelstein effect**

To elucidate the contributions of total nonequilibrium out-of-plane magnetization from orbit and spin moments in WTe$_2$, we calculated the current-induced orbital and spin magnetization. The spin magnetization $\boldsymbol{M}_{\text{spin}}$ is analogous to $\boldsymbol{M}_{\text{orb}}$ and can be described as $\boldsymbol{M}_{\text{spin}} = \left(\frac{e\tau}{\hbar}\right) \alpha_{\text{spin}} \boldsymbol{E}$. The spin Edelstein coefficient is evaluated as [57]

$$\alpha_{xz}^{\text{spin}} = \int_{BZ} \frac{d^2k}{(2\pi)^2} \sum_n m_{n\mathbf{k}}^{z,\text{spin}} v_{n\mathbf{k}}^x \partial_{\epsilon_{n\mathbf{k}}} f_{n\mathbf{k}}^{(0)}, \quad (S5)$$

where the spin magnetic moment $\boldsymbol{m}_{n\mathbf{k}}^{\text{spin}} = -\left\langle \partial_{\mathbf{k}} u_{n\mathbf{k}} \left| \frac{1}{2} g\mu_b \boldsymbol{\sigma} \right| \partial_{\mathbf{k}} u_{n\mathbf{k}} \right\rangle$, where $g \approx 2$ is the spin g-factor of the electron, $\boldsymbol{\sigma}$ is the Pauli operator and $\mu_b$ is the Bohr magneton.

**Supplemental Note 8: Experimental and calculated carrier densities of few-layer WTe$_2$.**

The semimetal WTe$_2$ is known to have a Fermi level changing markedly with temperature, which in turn leads to large variations in the electron and hole densities [59]. In order define the position of the Fermi level of WTe$_2$ sample, we conducted magnetotransport measurements at various temperatures to extract the carrier densities of few-layer WTe$_2$. Figures S6 shows the longitudinal ($\rho_{xx}$) and transverse ($\rho_{xy}$) resistivity of a few-layer WTe$_2$ (~5nm) device under the out-of-plane magnetic field at temperatures ranging from 1.6 K to 200 K. As temperature increases, the magnetoresistance effect rapidly becomes suppressed, and the Hall resistance transitions from a nonlinear to a linear dependence on magnetic field, indicating a transition from a two-carrier type to a single carrier type. The transition temperature is about 150 K in this device, which is consisted with the temperature-induced Lifshitz transition in WTe$_2$ (associated with the complete disappearance of the hole pockets) at $T \simeq 160$ K identified by ARPES [59]. Thus, we used a semiclassical two-carrier model [60] to extract the experimental hole and electron densities below 150 K and single-carrier model to extract the experimentally electron densities above 150K, respectively. Figure S7(a) shows the obtained temperature-dependent electron ($n_e$) and hole ($n_h$) densities. At 1.6 K, the hole and electron densities are nearly compensated. As the temperature increases from 1.6 to 100 K, the electron density increases, while the hole density drops by almost one order of magnitude.

To quantify the density variation from a theoretical perspective, we calculated the electron ($n_e$) and hole ($n_h$) densities as a function of chemical potential ($\mu$) from band-resolved density of states (Fig. S7(b)). The carrier densities are calculated by

$$n_e = \int_{\epsilon_c}^{+\infty} g_e(E) f_0(E - \mu) dE \tag{S6}$$

$$n_h = \int_{-\infty}^{\epsilon_v} g_h(E) f_0(E - \mu) dE \tag{S7}$$

where $g_e$ ($g_h$) is the density of states of electron (hole), $f_0$ is the Fermi-Dirac distribution function, $\epsilon_c$ ($\epsilon_v$) is the energy of conduction band minimum (valence band maximum). The Fermi energy of $WTe_2$ gained from static self-consistent calculation is set to $\mu = 0$ eV, corresponding to the chemical potential of the system in its intrinsic ground state at 0 K. At $\mu = 6$ meV, denoted by the grey dash line in Fig. S7(b), the calculated electron and hole densities are fully compensated. As the chemical potential increases, the electron density gradually increases while the hole density decreases rapidly. The calculated results agree with previous work [59].

To verify the corresponding Fermi level at different temperatures, using the method in previous works [26,65], we compared the experimental and calculated carrier densities of holes and electrons. Although the absolute value of calculated $n_h$ or $n_e$ shows some discrepancy compared to the experimental results, the similar trend as the temperature-dependence experimental densities can be captured by the calculated values versus the chemical potential $\mu$. At $T = 1.6$ K, where the experimental hole and electron densities are nearly compensated, this approximately corresponds to $\mu = 10$ meV in our first-principles calculations. Furthermore, when the temperature increases from 1.6 K to 100 K, the experimental electron density increases by about two times, while the hole density decreases by nearly an order of magnitude. The magnitude of these variations can be captured by the calculated values when $\mu$ varies from about 10 to 63 meV. We note that this effective shift in $\mu$ aligns with previous APRES study, which found that a temperature variation of 120 K induces a ~50 meV shift in the Fermi level of $WTe_2$ [59].

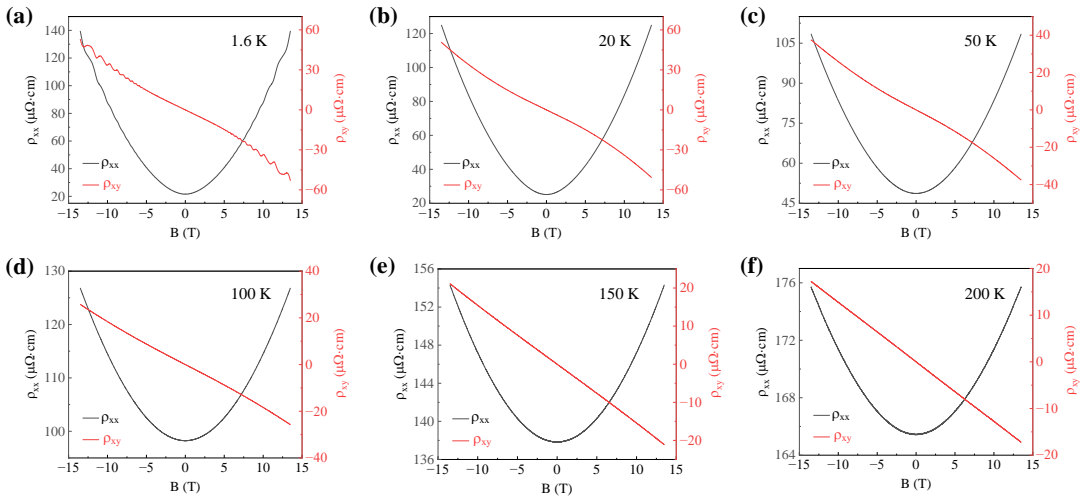

**FIG. S6.** Longitudinal and Hall resistivities as a function of the out-of-plane magnetic field at temperatures ranging from 1.6 K to 200 K.

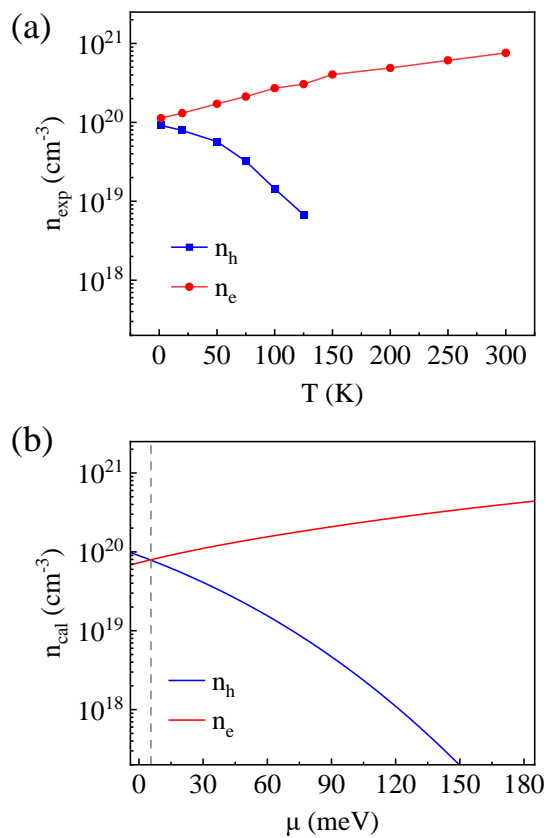

**FIG. S7. (a)** Experimental results on temperature dependence of the electron ($n_e$) and hole ($n_h$) carrier densities.
**(b)** Theoretically calculated electron ($n_e$) and hole ($n_h$) carrier densities versus the Fermi level $\mu$ of WTe$_2$.

**Supplemental Note 9: Circuit schematic with multiple sources in experiments.**

The circuit schematic in experiments (Fig. 4 in the main text) is depicted in Fig. S8. The AC, DC and pulsed DC sources are all effective current sources. The original SR830 AC source is a voltage source. In experiments, we connected the SR830 voltage source and a protective resistor with resistance value $R_{p1}$ of 1 MΩ. The resistance of WTe$_2$/FGT channel is on the order of 100 Ω, much less than $R_{p1}$. It makes the SR830 source an effective current source with excitation current $I^\omega = U^\omega/R_{p1}$, where $U^\omega$ is the source voltage (set to 1 V in experiments). The Keithley 2400 current source is used for the DC source. As shown in Fig. S8, the positive and negative terminals of the Keithley source are connected to a pair of diagonal electrodes along the $a$-axis of WTe$_2$ to form a loop circuit, i.e., a floating loop. The Keithley 6221 current source is used for applying the pulsed DC current with a duration of 60 μs. The impedance of the floating Keithley source to ground is measured to be ~60 MΩ. While, the negative terminal of SR830 and Keithley 6221 source is directly connected to the ground.

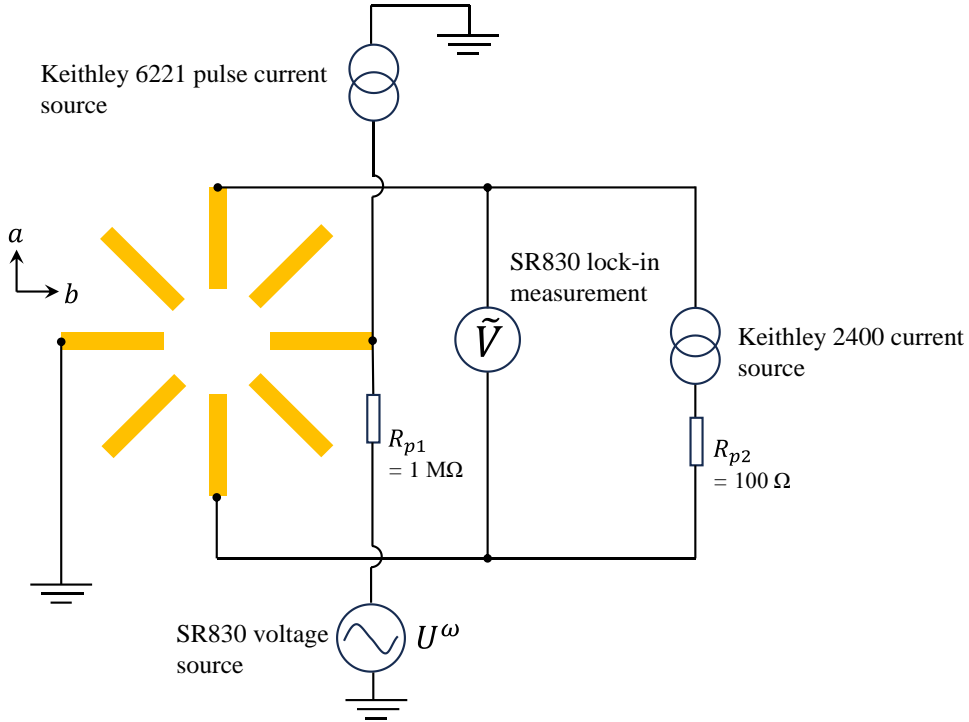

**FIG. S8.** The schematic of circuit structure utilized for measurements with two distinct driving currents, a pulsed current along the $b$-axis and a DC current along the $a$-axis of WTe$_2$. Besides, an AC current is applied for monitoring the anomalous Hall resistance.

**Supplemental Note 10: Reproducible results in Device C.**

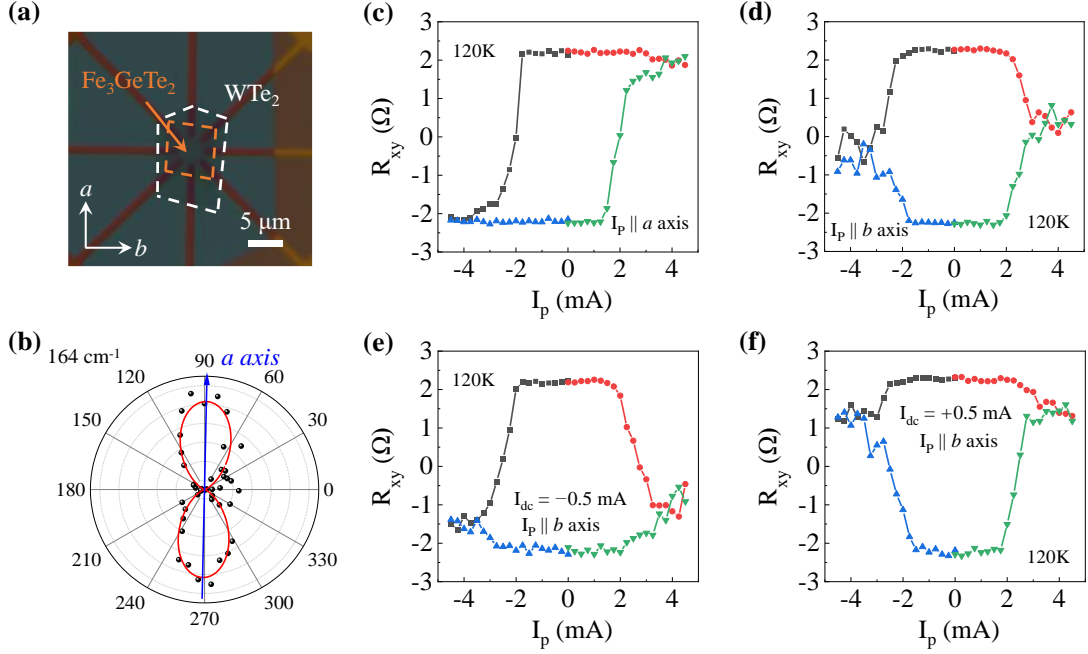

**FIG. S9. (a)** Optical image of Device C. Fe$_3$GeTe$_2$ and WTe$_2$ are distinctly delineated by orange and white dashed borders, respectively.

**(b)** Polarized Raman spectroscopy to identify the *a*-axis of WTe$_2$.

**(c, d)** Pulsed current $I_p$ induced field-free perpendicular magnetization switching at 120 K with $I_p$ applied along the (c) *a*-axis and (d) *b*-axis.

**(e, f)** The magnetization switching with two distinct driving currents for a pulsed current $I_p$ applied along the *b*-axis and $I_{dc}$ along the *a*-axis with (e) $I_{dc} = -0.5$ mA and (f) +0.5 mA at 120 K.

**Supplemental Note 11: AHE hysteresis loop shift measurements in Device A.**

We conducted anomalous Hall effect (AHE) hysteresis loop shift measurements to quantify the out-of-plane anti-damping-like torque efficiency. An out-of-plane anti-damping-like torque typically induces a shift in the AHE hysteresis loop to counteract intrinsic anti-damping once the current surpasses a critical threshold [61,62]. As depicted in Fig. S10(a), obvious AHE loop shifts are observed when applying positive (+3.5 mA) and negative (−3.5 mA) pulsed currents along the *a*-axis. In contrast, no significant loop shift is observed when $I_p$ (±3.5 mA) is applied along the *b*-axis. Note that the AHE loop shifts in Fig. S10(a) are larger than Fig. 4(f) and Fig. S10(c) owing to a larger generated out-of-plane anti-damping-like torque. Through extracting the loop shift field $H_{\text{shift}}$, we could calculate the out-of-plane anti-damping-like torque efficiency ($\xi_{\text{DL}}^z$) with the formula [63]

$$\frac{H_{\text{shift}}}{J_c} = \frac{\pi}{2} \frac{\hbar \xi_{\text{DL}}^z}{2e\mu_0 M_s t_{FGT}}, \tag{S8}$$

where $J_c$ is the critical current density, $\hbar$ is the reduced Planck's constant, $\mu_0$ is the vacuum permeability, $M_s$ is the saturation magnetization of Fe$_3$GeTe$_2$, and $t_{FGT}$ is the thickness of the FGT layer in our device. With $H_{\text{shift}}/J_c$ = 12 mT/(9.9×10$^6$ A/cm$^2$), $t_{FGT}$ = 6 nm, and the reported $M_s$ = 16 emu/cm$^3$ [64], we obtained the out-of-plane torque efficiency of 0.024 at 90 K for current along the *a*-axis.

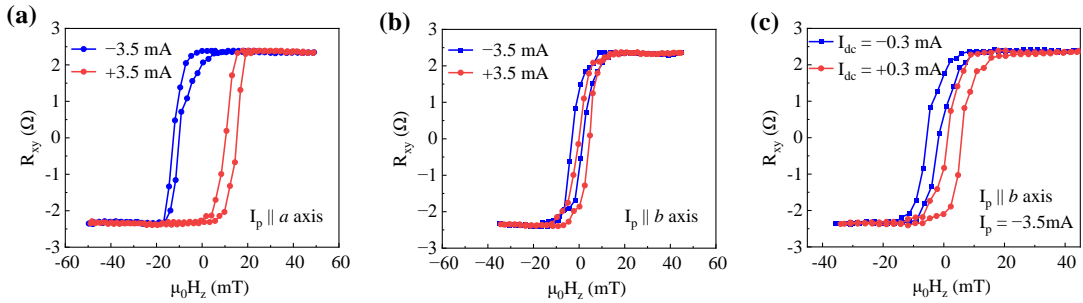

**FIG. S10. (a, b)** AHE hysteresis loops measured at 90 K with $I_p$ = ±3.5 mA along the (a) *a*-axis and (b) *b*-axis.
**(c)** AHE hysteresis loops measured at 90 K with $I_p$ = −3.5 mA along the *b*-axis and $I_{dc}$ = ±0.3 mA along the *a*-axis.

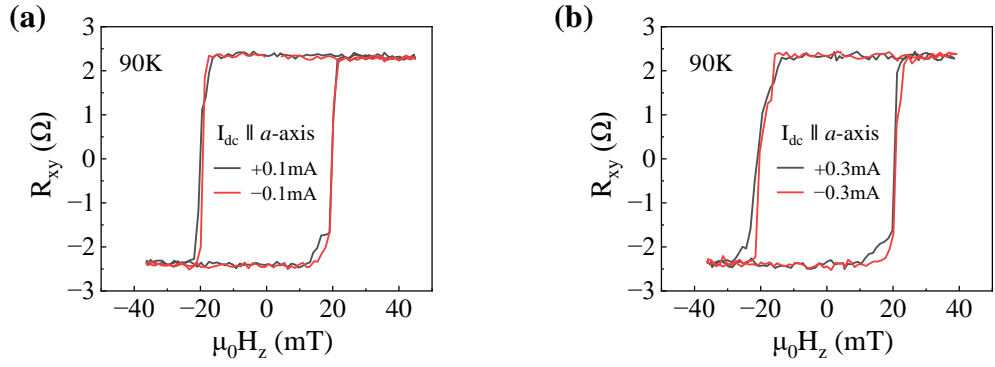

**FIG. S11.** AHE hysteresis loops measured at 90 K with (a) $I_{dc}$ = ±0.1 mA and (b) $I_{dc}$ = ±0.3 mA along the *a*-axis. There is no AHE loop shift in the absence of the pulsed current.

**Supplemental Note 12: Discussion of the heating effect in field-free perpendicular magnetization switching.**

An applied large current can generate Joule heating and increase the temperature of our device. However, although the coercive field of FGT is reduced owing to the Joule heating, it is clear from Fig. S10 and Fig. 4(f) that the temperature of the device remains below the Curie temperature in our experiments when a large pulsed current (surpasses the critical current) is applied.

Further, it should be recognized that the heating effect can contribute to magnetization switching. The heating effect can lower the coercive field and the energy barrier between the two magnetization states of FGT, which assists in current-induced magnetization switching. However, it's worth noting that the heating effect is not the result for achieving deterministic magnetization switching. Firstly, the observed current-induced magnetization switching in WTe$_2$/FGT heterostructure exhibits a strong dependence on the crystal axis angle of WTe$_2$, which is consistent with the mechanism of current-induced orbital magnetization. Secondly, upon comparison of the results shown in Fig. 4, it is evident that while the amplitudes of the applied currents are the same, indicating equal heating effects, the direction of magnetization switching is clearly determined by the sign of the DC current along the *a*-axis.

Therefore, the heating effect may assist the magnetization switching but cannot play a deterministic role on the polarity of the field-free magnetization switching.

**Supplemental Note 13: Micromagnetic simulations.**

To further understand the collaborative dynamics between the out-of-plane and in-plane anti-damping-like torques in magnetization switching, micromagnetic simulations are performed using Mumax$^3$ [51] for a multi-domains model of FGT. The magnetization switching dynamics is governed by the Landau-Lifshitz-Gilbert (LLG) equation,

$$\frac{dm}{dt} = -\gamma\mu_0 m \times (H_{\text{eff}} + H_{\text{ext}}) + \alpha m \times \frac{dm}{dt} + \tau_{\text{SOT}}, \quad (S9)$$

where $\tau_{\text{SOT}}$ contains the field-like (FL) and anti-damping like (DL) torque. In Mumax$^3$, the spin torques are characterized by a Slonczewski STT model with a fixed layer and a free layer. The Slonczewski STT model is equivalent to the SOT model by setting the spacer layer thickness $\Lambda = 1$, the secondary spin-torque parameter $\varepsilon' = \eta\varepsilon$ and the spin polarization $P = \theta_{\text{DL}}$, where $\eta$ is the ratio of FL and DL torques, and $\theta_{\text{DL}}$ is the charge-to-spin conversion efficiency. Additionally, the tilted angle of injected spins is defined by setting the canting angle of the fixed layer's magnetization.

The simulation grid in our simulations is defined as 512 × 512 × 1 with a cell size of 2 × 2 × 1 nm. The simulation parameters are chosen to closely reflect experimental results: the saturation magnetization $M_s$ = 1.7 ×10$^5$ A/m, the exchange constant $A_{\text{ex}}$ = 10 pJ/m, the uniaxial anisotropy constant $K_u$ = 5.1 × 10$^5$ J/m$^3$, and the Gilbert damping constant α = 0.02. The ratio of FL and DL torques is set as 0.02 and the charge-to-spin conversion efficiency is set as 0.2.

Moreover, a spin polarization $\boldsymbol{\sigma} = (\cos\theta, 0, \sin\theta)$ of the fixed layer is set, where $\theta$ is the canting angle and thus $\tan\theta$ represents the ratio of the out-of-plane and in-plane torque components. According to the torque efficiencies of WTe$_2$ obtained in our experiments and reported studies, $\theta = 0°$, $\theta = -1.09°$, and $\theta = -12°$ approximately correspond to the cases of $I_p$ along the *b*-axis alone, $I_p$ along the *b*-axis with a $I_{dc}$ = +0.4 mA along the *a*-axis, and $I_p$ along the *a*-axis alone in our experiments, respectively. By varying the angle $\theta$, magnetization reversal processes with different ratios of the out-of-plane and in-plane anti-damping-like torques are investigated.

Figure S12(a) shows the time evolution of magnetization switching under a 1 ns pulsed current with a current density of $7.5 \times 10^8$ A/cm$^2$ for different angles $\theta$. Note that the applied current density surpasses the threshold current density (Fig. S13). The magnetization is initially driven into an in-plane direction and remains in this state during the pulse duration, subsequently relaxing to a steady state after the pulse is removed. Moreover, we identify the stable magnetization at 10 ns as the final magnetization $M_z^{Final}$. The magnetization switching is strongly influenced by $\theta$. In the absence of the out-of-plane anti-damping-like torque ($\theta = 0°$), the upward and downward domains are equally distributed in FGT [Fig. S14(c)], resulting in a nearly zero value of $M_z^{Final}$. As $\theta$ increases from 0° to −2°, partial magnetization switching is observed, with a larger magnitude of $M_z^{Final}$ observed for larger $\theta$. When $\theta$ reaches −2°, full magnetization switching is achieved. Figure S12(b) displays the obtained $M_z^{Final}$ as a function of $\theta$ in simulations, which is well consistent with the results obtained in our experiments [Fig. 4(e)]. To visually understand the switching process, the spatially resolved domain evolutions with $\theta = +1.09°$ and $\theta = −1.09°$ are displayed in Fig. S12(c) and Fig. S12(d), respectively. The switching process is accompanied by the nucleation of multiple domains and the propagation of domain walls. For $\theta = +1.09°$, the multidomain pattern stabilizes at 10 ns, with a larger region occupied by upward-oriented magnetic domains [Fig. S12(c)], resulting in net magnetization state along the +z direction. Conversely, for $\theta = −1.09°$, the situation is reversed [Fig. S12(d)].

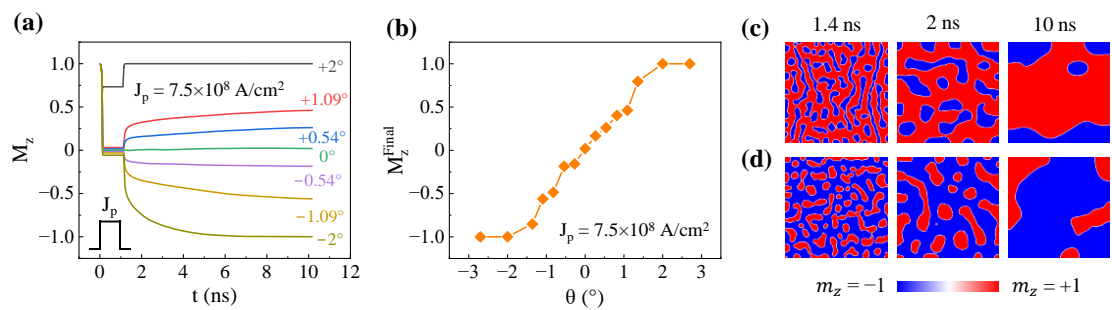

**FIG. S12. (a)** Magnetization switching for different angles $\theta$. The magnetization state is initialized along the +z direction. A pulsed current $J_p = 7.5 \times 10^8$ A/cm$^2$ with a duration of 1 ns is applied in simulations.

**(b)** The final net magnetization ($M_z^{final}$) calculated as a function of angle $\theta$ under the pulse current $J_p = 7.5 \times 10^8$ A/cm$^2$.

**(c, d)** Snapshots recorded at distinct times of the out-of-plane component of magnetic moments $m_z$ during the switching process under a 1 ns pulsed current $J_p = 7.5 \times 10^8$ A/cm$^2$ for (c) $\theta = +1.09°$ (d) and $-1.09°$, respectively. The red and blue colors indicate that the magnetic moments point along +z and –z, respectively, whereas the white color represents the in-plane magnetic moments.

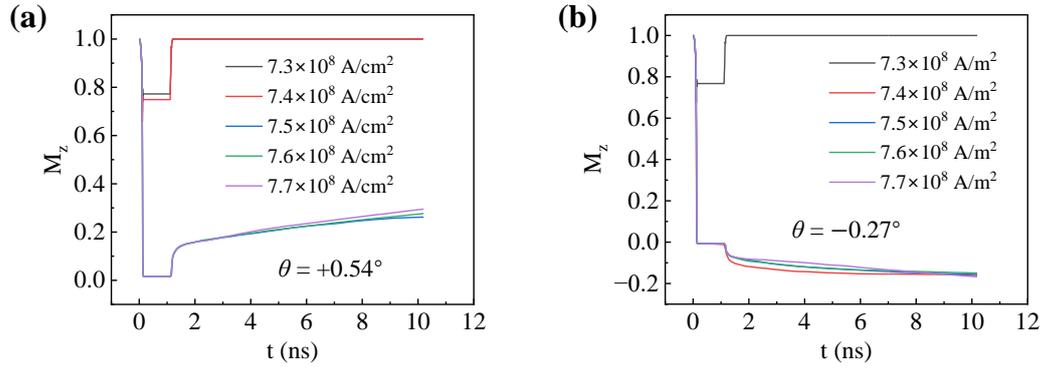

**FIG. S13. (a, b)** Time-resolved magnetization switching for different current densities with angles of $\theta$ of (a) $+0.54°$ and (b) $-0.27°$, respectively. The magnetization state is initialized in the +z direction. The duration of pulsed currents is 1 ns.

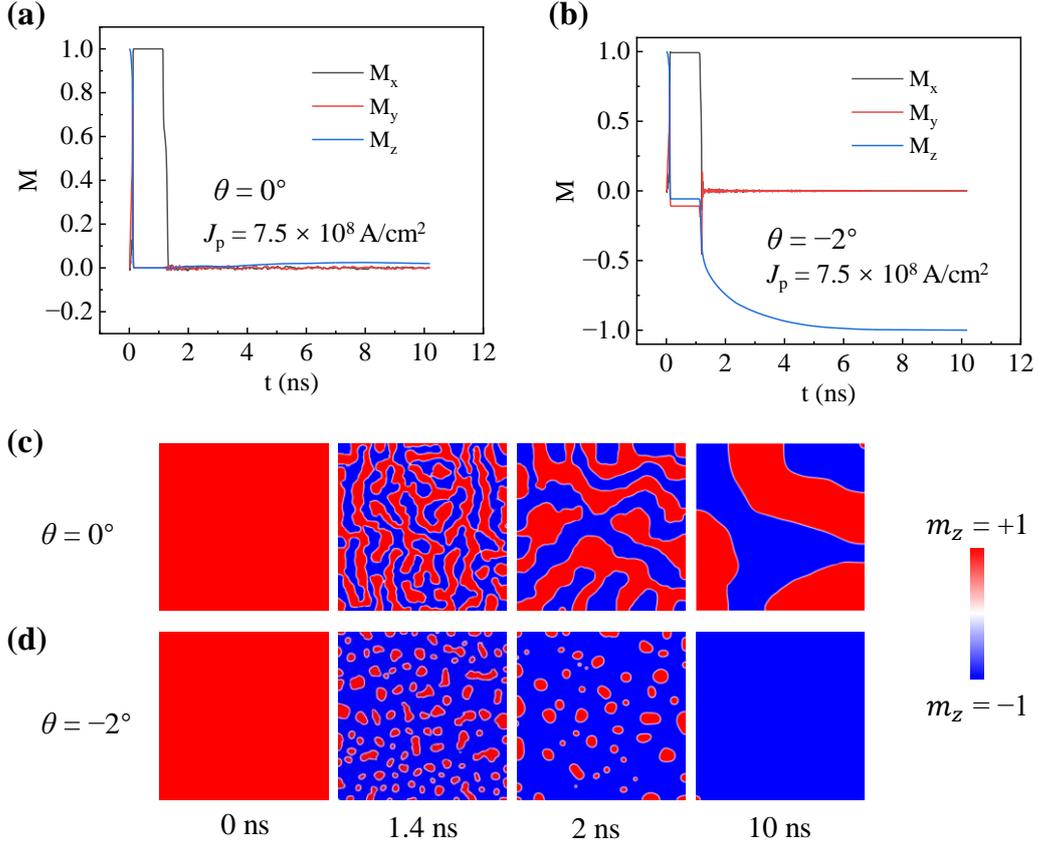

**FIG. S14. (a, b)** Time-resolved magnetization switching under a 1-ns pulsed current with a current density of $7.5 \times 10^8$ A/cm$^2$ for angles of (a) 0° and (b) −2°.

**(c, d)** Snapshots recorded at distinct times of the $z$ component of magnetic moments $m_z$ during the switching process under a 1 ns pulsed current $J_p = 7.5 \times 10^8$ A/cm$^2$ for (c) $\theta = 0°$ and (d) –2°. The red and blue colors indicate that the magnetic moments point along +z and –z, respectively, whereas the white color represents the in-plane alignment of magnetic moments.